\documentclass[10pt,journal]{IEEEtran}

\usepackage{amsfonts}
\usepackage{amsmath}
\usepackage{amssymb}
\usepackage{amsthm}
\usepackage{calc}

\usepackage{enumitem}
\usepackage{etoolbox}
\usepackage{graphicx}
\usepackage{mathtools}
\usepackage{subcaption}
\usepackage{floatrow}
\usepackage{booktabs}

\newcommand{\definition}[1]{\textbf{#1}}

\DeclareMathOperator*{\ARGMIN}{argmin}
\newcommand{\argmin}{\ARGMIN\limits}
\DeclareMathOperator*{\spath}{sp}
\DeclareMathOperator*{\st}{s.t.}
\DeclareMathOperator*{\rank}{rank}

\makeatletter
\g@addto@macro\th@plain{\thm@headpunct{\ }}
\makeatother
\makeatletter
\g@addto@macro\th@definition{\thm@headpunct{\ }}
\makeatother
\csundef{proof}
\csundef{endproof}
\theoremstyle{plain}\newtheorem{theorem}{Theorem}
\theoremstyle{plain}
\theoremstyle{plain}
\theoremstyle{plain}\newtheorem{remark}{Remark}
\theoremstyle{definition}\newtheorem*{proof}{Proof}
\theoremstyle{definition}

\begin{document}

	%***Paper heading data***
	
	%***Title***
	\title{Structurally Observable Distributed Networks of Agents under Cost and Robustness Constraints}
	%***Author***
	\author{Stephen~Kruzick,~%\IEEEmembership{???,~IEEE,}
        S\'{e}rgio~Pequito,~%\IEEEmembership{???,~IEEE,}
        Soummya~Kar,~%\IEEEmembership{???,~IEEE,}\\
        Jos\'{e}~M.~F.~Moura,~%\IEEEmembership{???,~IEEE,}
        and~A.~Pedro~Aguiar%,~\IEEEmembership{???,~IEEE}%
			\thanks{Stephen~Kruzick (skruzick@andrew.cmu.edu), S\'{e}rgio~Pequito (sergo@seas.upenn.edu), Soummya~Kar (soummyak@andrew.cmu.edu), and Jos\'{e}~M.~F.~Moura (moura@ece.cmu.edu) affiliate with the Department of Electrical and Computer Engineering at Carnegie Mellon University in Pittsburgh, PA, USA.  A.~Pedro~Aguiar (pedro.aguiar@fe.up.pt) affiliates with the Department of Electrical and Computer Engineering at University of Porto in Porto, Portugal.  The work of Kruzick was partially conducted with government support under and awarded by the DoD, Air Force Office of Scientific Research, National Defense Science and Engineering Graduate (NDSEG) Fellowship, 32 CFR 168a and by NSF grants CCF-1018509, CCF-1011903, and CCF-1513936.  The work of Pequito was partially supported by grant SFRH/BD/33779/2009 from Funda\c{c}\~{a}o para a Ci\^{e}ncia e a Tecnologia (FCT)  and by the CMU-Portugal (ICTI) program.}}
			
%***Generate title section***
	\maketitle
	
	%***Abstract section***
%	\IEEEtitleabstractindextext{
	\begin{abstract}\label{Abstract}
In many problems, agents cooperate locally so that a leader or fusion center can infer the state of every agent from probing the state of only a small number of agents. Versions of this problem arise when a fusion center reconstructs an extended physical field by accessing the state of just a few of the sensors measuring the field, or a leader monitors the formation of a team of robots. Given a link cost, the paper presents a polynomial time algorithm to design a minimum cost coordinated network dynamics followed by the agents, under an observability constraint. The problem is placed in the context of structural observability and solved even when up to~$k$ agents in the coordinated network dynamics fail.
	\end{abstract}
	
	%***Keywords section***
	\begin{IEEEkeywords}\label{Keywords}
		distributed networks, linear dynamics, structural systems, observability, minimum cost, Steiner subgraph
	\end{IEEEkeywords}
	%}

	%***Introduction section***
	\section{Introduction}\label{Introduction}
		%\IEEEPARstart{E}{xtensive} study has been conducted regarding networks operating as linear dynamic systems to produce a global desired behavior from simple local interactions.  Typically, each node in the network is capable of transmitting stored data to neighboring nodes and receiving data transmitted by neighboring nodes, as well as possibly making measurements or taking actions.  Alternatively, in some network systems, states are not actively transmitted to neighboring nodes, but are instead measured or sensed by the receiving node.  This can be found in the natural world.  For instance, flocks of birds or schools of fish can be described by models in which each individual constantly monitors or senses the distance to neighboring formation members and reacts according to some dynamics to ensure that it maintains a suitable position.  The qualitative properties of such dynamics are discussed in \cite{??}, \cite{??}, \cite{??} and \cite{??} along with examples.  These dynamics are examined in the context of consensus algorithms in \cite{??}, and the self-organization of these adaptive networks has also been studied in the signal processing literature in \cite{??}.
\IEEEPARstart{I}{t} is important in many distributed settings to appreciate the impact of local cooperation. Much work has been done in distributed inference \cite{bracamaranomatta-08,karmouraramanan-tit-2012}, distributed diffusion \cite{lopessayed-08}, and distributed optimization \cite{nedicozdaglar-09,jakoveticxaviermoura-tac-14}, to name a few applications and references. This paper is concerned with a design question under the following different but related scenario. A group of agents share a common goal. They cooperate at a cost to update their individual state, but only a few agents share their state with a leader or fusion center.  This leads to network dynamics that we will describe more formally below. The mission of the leader is to reconstruct the network state, i.e., the state of all the agents. We consider how to design the minimal cost network of (local) cooperation among agents and with which agents the leader interacts, i.e., the design of minimal cost coordinated network dynamics, subject to the constraint that the leader can reconstruct the network state--an observability constraint. Before being more specific, we illustrate with two motivating examples. The first is static, while the second is dynamic like the problem we address. References \cite{schmidtmoura2,schmidtmoura3} consider the problem of reconstructing a field from information (state) provided by only a few agents (sensors) in a network of sensors. For instance, the field may be the contamination level of a pollutant over a wide area. It turns out that, through local cooperation (sensors updating iteratively their state with the state of their neighbors), the fusion center can, under appropriate conditions, reconstruct (through a basis pursuit type algorithm) the entire field from a snapshot of the states of a very sparse subset of the agents. A question of interest is to design the cooperation dynamics and to determine a minimal set of sensors to be probed so that a fusion center can reconstruct the field.

The second considers formation with flocks of birds or schools of fish that can be described by models in which each individual constantly monitors or senses the distance to neighboring birds or fish and reacts to ensure that it maintains a suitable position. For example, each individual might accelerate to move closer if the distance to neighbors increases or slow down if the distance to neighbors becomes dangerously close. The collective networked dynamics of birds or fish and their qualitative properties are discussed in \cite{reynolds-87,VicsekCzirokBenJacobShochet-95,CuckerSmale-07}, among others. These references provide several interesting and revealing examples.  The self-organization of these adaptive networks has also been studied  in~\cite{tusayed-08}. Similarly, in several robotic problems like in formation control of multi-robot teams and in the coordination of groups of mobile autonomous agents,
%~\cite{JadbabaieLinMorse-03}, which cast these problems in the consensus algorithm framework, see also \cite{DistAvgCons,boyd2006randomized,GossipAlg},
 robots move by approaching or gaining distance from their neighbors in order to maintain the formation.  This leads the multi-robot team to follow coordinated network dynamics. Furthermore, the robots do not explicitly communicate with each other in many of these applications. Each robot finds its own relevant information with respect to neighboring robots~\cite{Moshtagh2009}, for example, range and bearing from on-board sensors, such as cameras.  However, for reasons derived from the application, the robots do not directly communicate this information to other individuals. The resulting networked dynamics have been used in the literature to obtain simplified linear distributed control or update laws achieving mission specific formation or flocking objectives~\cite{cao2011formation,sun2015rigid}. Finally, we assume that a leader is charged with monitoring that the robots remain in formation by probing only the state of a few agents and then computing the (global) network state. This task is trivial if the fusion center accesses every agent or node in the network.  It becomes interesting when at the same time it is desirable to reduce the costs to a certain minimum, see below, including agents measuring their relative position to neighbors and the direct access to node states by the leader. It is then of interest to design the network structure of the coordinated dynamics and to determine which node states should be measured by the leader so that it can track the states of all the nodes in the network while minimizing the cost of network coordination and measurements.

The above are difficult questions. We cast problems of these types below in the context of structural observability, which entails designing the structure of a dynamical system with observable dynamics, a framework studied in~\cite{JiEgerstedt,SundaramHadjicostis1,SundaramHadjicostis2}.  Once the system is observable, the actual problem of continuously monitoring the state of the nodes by the remote fusion station can then be achieved by implementing a recursive observer such as, for example, of the Luenberger type~\cite{ellis2002observers}.
  \begin{remark}
  We emphasize that there may not be explicit communication among the nodes in the applications we envision, even though we use the consensus context to describe the coordination among agents.  Accordingly, in this paper, we replace communication costs usually assumed in consensus problems by link costs. Link costs subsume costs associated with local neighbor sensing interactions (proxies for costs associated with sensing and extracting from measurements the state of neighbors), as well as costs incurred by a fusion center to learn (also, possibly by sensing) the current state (position) of a few selected agents. For simplicity, the interactions between the fusion center and nodes are described as through a \textit{backbone} network.
 %\hfill $\small \blacksquare$
 \end{remark}

		We model the coordinated network dynamics followed by the aggregate of all the agents by a linear system
		\begin{alignat}{4}
			&\mathbf{x}({n+1})&&=A\mathbf{x}(n)\label{dyn1.1} \\
			&\mathbf{y}({n}) &&=C\mathbf{x}(n)\label{dyn1.2}.
		\end{alignat}
        The system dynamics matrix~$A$ captures the network graph of interactions among agents.  Its nonzero entries represent interaction links between corresponding nodes or agents. Each agent or sensor node maintains a single scalar state variable $\mathbf{x}_i(n)$ initially set to $\mathbf{x}_i(0)$.  The vector of agent or sensor states at a given iteration $n$ is denoted by $\mathbf{x}(n)$.  The nodes update their states according to the coordinated dynamics given by~\eqref{dyn1.1}, with initial state vector $\mathbf{x}(0)$.  The nodes that are monitored by the fusion center, referred to as backbone nodes, are collected in the output vector $\mathbf{y}(n)$ given by~\eqref{dyn1.2}.  Because network structure must be respected, disallowed interactions among nodes or among the fusion center and nodes restrict corresponding entries of the $A$ and $C$ matrices to equal zero, whereas entries corresponding to network links represent design parameters.

		We use structural systems theory, a survey of which can be found in~\cite{StructSysSurvey}, to find network dynamics $(A,C)$ requiring a minimum set of link costs while guaranteeing that the initial state can be recovered from the backbone outputs collected over time at the fusion center despite some number of sensor node failures, which are known to the central node and occur before the dynamics begin.  Networks that operate according to this framework are discussed in~\cite{JiEgerstedt}. This reference provides a necessary topological condition for the initial state to be reconstructed from the backbone outputs for a specific choice of dynamics related to the network graph structure. These networks are also studied under a structural systems context in~\cite{SundaramHadjicostis1} and~\cite{SundaramHadjicostis2}, which describe types of networks where this can be achieved.  Our paper focus on designing robust networks that operate according to the above model and that minimize a specified cost function.  This work extends a preliminary version of the optimal network dynamics design problem that we previously introduced in~\cite{PequitoKruzickEUSIPCO2013}, by considering 
arbitrary costs, arbitrary backbone topologies, and sensing robustness requirements. From a technical standpoint, the methodology and combinatorial optimization tools are also more general and computationally efficient. Practical implementable dynamic systems of the optimal network structures that we obtain rely on results presented in~\cite{SundaramHadjicostis2}.  While the optimization problem involving cost minimization under robustness constraints that appears in this paper does not have a direct comparison in the existing structural systems literature, other related works regarding structural systems may be of interest for further reading.  The structural systems framework for networks appears in \cite{JiEgerstedt}, \cite{SundaramHadjicostis1}, and \cite{SundaramHadjicostis2}.  Other optimal network design problems involving this structural systems framework with significantly differing objectives and constraints can be found in \cite{SergioA}, \cite{SergioB}, and~\cite{SergioD}.

%Other optimal designs of structural systems not directly related to ours are in~\cite{SergioA}, \cite{SergioB}, and~\cite{SergioD}.

		%***Paper content preview***
		The paper is organized as follows.  Section~\ref{BackgroundConcepts} introduces key background concepts, including relevant information from graph theory, combinatorial optimization, and systems theory.  Section~\ref{NetworkDescription} formally describes our network operation model.  Section~\ref{DesignProblem} examines our minimum cost design problem and provides a solution algorithm followed by a proof of correctness and practical discussion.  Finally, Section~\ref{Conclusions} concludes the paper.

	%***Background section***
	\section{Background Concepts}\label{BackgroundConcepts}
	
		%***Section introduction***	
		This section provides supporting background information.  Of particular importance are graph theory definitions and concepts used to describe network structure as well as topics from combinatorial optimization relevant to the minimum cost design problem solutions.  Network operation closely relates to system theory concepts, which are also introduced.  Finally, results from structural system theory provide a bridge between network structure, described by graphs, and network operation, described by dynamical systems.	
		
		%***Graph theory definitions and concepts***
		\subsection{Graph Theory Concepts}\label{BackgroundConcepts:GraphTheory}
		
			%***Basic definitions***
			This section introduces terminology and concepts regarding graphs.  A \definition{directed graph} $\mathcal{G}$ is an ordered pair $(\mathcal{V},\mathcal{E})$ in which $\mathcal{V}$ denotes a set of \definition{nodes (vertices)} and $\mathcal{E}$ denotes a set of directed \definition{links (edges)}.  These directed links are ordered pairs $(v_i,v_j)$ of nodes $v_i,v_j\in \mathcal{V}$.  Note that \definition{self-loops}, links formed as $\left(v,v\right)$ for $v\in \mathcal{V}$, are not excluded from this definition.  Furthermore, an \definition{directed, weighted graph} $\mathcal{G}$ is the ordered triple  $(\mathcal{V},\mathcal{E},w)$ in which $w:\mathcal{E}\rightarrow\mathbb{R}^+$ assigns a cost or \definition{weight} $w(e)$ to each link $e\in \mathcal{E}$.  Any graph $\mathcal{G}_S=(\mathcal{V}_S,\mathcal{E}_S)$ with $\mathcal{V}_S\subseteq \mathcal{V}$ and $\mathcal{E}_S\subseteq \mathcal{E}$ is called a \definition{subgraph} of $\mathcal{G}$.  Furthermore, if $\mathcal{V}_S=\mathcal{V}$, then $\mathcal{G}_S$ \definition{spans} $\mathcal{G}$.  Two graphs $\mathcal{G}_1=(\mathcal{V}_1,\mathcal{E}_1)$ and $\mathcal{G}_2=(\mathcal{V}_2,\mathcal{E}_2)$ are said to be \definition{isomorphic}, written as $\mathcal{G}_1\simeq \mathcal{G}_2$, if there is bijective function $f:\mathcal{V}_1\rightarrow \mathcal{V}_2$ such that $(u,v)\in \mathcal{E}_1$ if and only if $(f(u),f(v))\in \mathcal{E}_2$.

			A sequence $(v_1,v_2),(v_2,v_3),...,(v_{k-1},v_k)$ of directed links in which the \definition{head} (destination) node of the previous link is the \definition{tail} (origin) node of the subsequent link constitutes a \definition{directed path}.  Provided $v_i\neq v_j$ for all $i\neq j$, it comprises an \definition{directed elementary path}.  When $v_1=v_k$ but all other nodes are distinct, the path forms a \definition{directed cycle}.   Two directed paths are \definition{internally node-disjoint} if they share no nodes apart from the start node and end node.  Likewise, two directed paths are \definition{link-disjoint} if they share no links.  The \definition{directed local node-connectivity} $\kappa_{\mathcal{G}}(u,v)$ from node~$u$ to node~$v$ gives the minimum number of nodes that must be removed from the graph $\mathcal{G}$ such that there is no directed path from~$u$ to~$v$, equal to the number of internally node-disjoint directed paths from~$u$ to~$v$.  The \definition{directed local link-connectivity} $\lambda_{\mathcal{G}}(u,v)$ from node~$u$ to node~$v$ gives the minimum number of links that must be removed from the graph such that there is no directed path from~$u$ to~$v$, equal to the number of link-disjoint directed paths from~$u$ to~$v$.			
			
			%A sequence $(v_1,v_2),(v_2,v_3),...,(v_{k-1},v_k)$ of links in which consecutive elements share a node constitutes a \definition{path}.  Provided $v_i\neq v_j$ for all $i\neq j$, it comprises an \definition{elementary path}.  When $v_1=v_k$ but all other nodes are distinct, the path forms a \definition{cycle}.   Two paths are \definition{internally node-disjoint} if they share no nodes apart from the start node and end node.  Likewise, two paths are \definition{link-disjoint} if they share no links.  The \definition{local node-connectivity} $\kappa_{\mathcal{G}}(u,v)$ between nodes~$u$ and~$v$ gives the minimum number of nodes that must be removed from the graph $\mathcal{G}$ such that there is no path from~$u$ to~$v$, equal to the number of internally node-disjoint paths from~$u$ to~$v$.  The \definition{local link-connectivity} $\lambda_{\mathcal{G}}(u,v)$ between nodes~$u$ and~$v$ gives the minimum number of links that must be removed from the graph such that there is no path from~$u$ to~$v$, equal to the number of link-disjoint paths from~$u$ to~$v$.  Directed versions of these concepts are readily defined when graphs with directed links are considered.
			
			%***Connectivity definitions***
			Several notions of connectedness exist for directed graphs.  In this paper, a specially labeled \definition{root} node $r\in \mathcal{V}$ is given, and a graph is \definition{$\mathbf{r}$-rooted connected} if there is an elementary directed path from each node $v\in \mathcal{V}$ to~$r$.  An \definition{arborescence}, also known as a directed rooted tree, is a directed graph in which there exists exactly one elementary directed path from each node to~$r$.  A collection of disjoint arborescences with root nodes collected into a set~$R$ is called a \definition{branching}.  A directed graph is \definition{$\mathbf{r}$-rooted $\mathbf{k}$-node-connected} if removing fewer than~$k$ nodes leaves at least one elementary directed path from each node to~$r$, that is $\kappa_{\mathcal{G}}(v,r)\geq k$ for all $v\in \mathcal{V}$.  Similarly, a directed graph is \definition{$\mathbf{r}$-rooted $\mathbf{k}$-link-connected} if removing fewer than~$k$ links leaves at least one elementary directed path from each node to~$r$, that is $\lambda_{\mathcal{G}}(v,r)\geq k$ for all $v\in \mathcal{V}$.			

		\subsection{Optimal Connectivity Problems}\label{BackgroundConcepts:OptimizationProblems}
		
			%***Overview***
			Network design problems often involve finding optimal subgraph structures within a graph of possible network connections with the restriction that some notion of connectivity be ensured.  The following discussion introduces problems involving optimality objectives and connectedness requirements that will prove useful in solving the problem presented in Section~\ref{DesignProblem}.

	Consider identification of the least costly set of directed links that connects all the nodes of a network to a root node.  The \definition{minimum spanning arborescence}, also known as a minimum directed rooted spanning tree, of a directed, weighted graph $\mathcal{G}=(\mathcal{V},\mathcal{E},w)$ rooted at a node $r\in \mathcal{V}$ formalizes this concept.  The problem consists of finding a subgraph of minimum cost $\operatorname{MSpA}(\mathcal{G},r)$ that has exactly one directed path from each node $v\in \mathcal{V}$ to~$r$.  Note that the solution must be an $r$-rooted arborescence and that $\mathcal{G}$ must be $r$-rooted connected for a solution to exist.  The minimum spanning arborescence can be computed by the Chu-Liu/Edmonds Algorithm, which can be implemented with complexity $O(|\mathcal{E}|+|\mathcal{V}|\log |\mathcal{V}|)$~\cite{Gabow}.  Generalizations for higher connectivity, such as the minimum $r$-rooted $k$-node-connected spanning subgraph and the minimum $r$-rooted $k$-link-connected spanning subgraph are similarly defined.  Unlike the corresponding undirected counterparts, these directed, rooted problems are solvable in polynomial time using a maximum weight matroid intersection formulation~\cite{Edmonds}, \cite{FrankRootedKConnections}.

In a problem related to the minimum spanning arborescence, consider identification of the least costly set of directed links that connect a required subset of the network nodes to a root node while possibly involving other nodes.  The general \definition{minimum Steiner arborescence} problem, also known as the minimum directed rooted Steiner tree, formalizes this concept.  Let a directed, weighted graph $\mathcal{G}=(\mathcal{V},\mathcal{E},w)$ with root node $r\in\mathcal{V}$ be given along with a set $\mathcal{S}\subseteq \mathcal{V}$ of \definition{terminal nodes}.  The problem consists of finding a minimum cost subgraph $\operatorname{MStA}(\mathcal{G},\mathcal{S},r)$ of~$\mathcal{G}$ that has exactly one directed path from each node $v\in \mathcal{S}$ to $r$.  Note that non-terminal nodes in $\mathcal{V}\backslash\mathcal{S}$ may or may not be used, and that a path must exist from each $v\in \mathcal{V}$ to $r$ for a solution rooted at~$r$ to exist.  The general minimum Steiner arborescence problem is NP-hard, so no polynomial time solution algorithm is known.  However, polynomial time approximation algorithms with nontrivial performance guarantee ratios exist and may be used to obtain approximate solutions~\cite{DirectedSteiner}.
			
		%***Minimum r-rooted k-connected Steiner arborescence***
		A generalization to greater connectivity requirements, the \definition{minimum $\mathbf{r}$-rooted $\mathbf{k}$-node-connected Steiner subgraph} problem, provides the minimum cost connectivity problem of greatest utility to this paper.  For a directed, weighted graph $\mathcal{G}=(\mathcal{V},\mathcal{E},w)$, terminal node set~$\mathcal{S}$, and root node~$r$, the problem consists of finding a minimum cost subgraph $\operatorname{MStNCS}(\mathcal{G},\mathcal{S},r,k)$ that has~$k$ internally node-disjoint directed paths from each node $v\in \mathcal{S}$ to~$r$.  Note that non-terminal nodes in $\mathcal{V}\backslash\mathcal{S}$ do not necessarily have~$k$ node-disjoint paths to~$r$.  It is clear that at least~$k$ internally node-disjoint paths must exist from each $v\in \mathcal{S}$ to~$r$ for a solution rooted at~$r$ to exist.  The problem is, in general, at least as computationally difficult as the minimum Steiner arborescence problem.  However, in the particular restricted case in which all links with positive weight, called \definition{augmenting links}, originate in~$\mathcal{S}$, the problem is solvable in polynomial time using a submodular flow algorithm \cite{FrankRootedKConnections}.

	For networks that engage in data forwarding, the problem of finding the path between two nodes with the least costly total link sum often finds relevance.  The concept of each node connecting to a root node over the least costly path possible is formalized by the \definition{shortest path spanning arborescence}, which will also contribute to solving the design problem in Section~\ref{DesignProblem}.  Let a directed, weighted, and $r$-rooted connected graph $\mathcal{G}=(\mathcal{V},\mathcal{E},w)$ with nonnegative link weights~$w$ and a root node $r\in \mathcal{V}$ be given.  The problem consists of finding a path from each $v\in \mathcal{V}$ to the root node~$r$ that has the minimum possible link sum.  The solution set of links is not necessarily unique.  However, at least one possible set of links forms a spanning tree called the shortest path spanning arborescence $\operatorname{SPSpA}(\mathcal{G},r)$ rooted at~$r$.  The shortest path lengths, using which the arborescence may be built starting at the root, can be computed using methods such as Dijkstra's algorithm, which can be implemented with complexity $O(|\mathcal{E}|+|\mathcal{V}|\log|\mathcal{V}|)$~\cite{CombinatorialOpt}.
		
		%***System theory definitions and concepts***
		\vspace{-10pt}
		\subsection{System Theory Concepts}\label{BackgroundConcepts:SystemTheory}
		
			%***LTI systems***
			The weighted networks discussed in this paper follow the coordinated dynamics  given by the discrete time dynamical system~\eqref{dyn1.1} and~\eqref{dyn1.2}.
 %A \definition{dynamical system} over a field $\mathbb{F}$ consists of a (vector) \definition{state} variable $\mathbf{x}(n)\in\mathbb{F}^{N}$ and a rule for computing subsequent values $\mathbf{x}({n+1})$ of the state variable from the existing state value $\mathbf{x}(n)$ and the (vector) system \definition{input} $\mathbf{u}(n)$.  A (vector) system \definition{output} $\mathbf{y}(n)$ may also be computed at each time from the concurrent state and input values.  Specifically, in the protocols to be described, the network coordination dynamics constitute a \definition{linear, time-invariant} systemnrepresented by
%		\begin{alignat}{4}
%			&\mathbf{x}({n+1}) &&= A \mathbf{x}(n) + B \mathbf{u}(n) \label{syst1} \\
%			&\mathbf{y}({n})   &&= C \mathbf{x}(n) + D \mathbf{u}(n) \label{syst2}.
%		\end{alignat}
%			
			%***Observability***
			The concept of system observability plays a key role in understanding the communication scheme under consideration.  A system is said to be \definition{observable} if the initial state $\mathbf{x}(0)$ can be uniquely determined from the collected outputs over a finite time period with knowledge of the system matrices and input values.  For discrete linear, time-invariant systems with zero-input such as in~\eqref{dyn1.1}-\eqref{dyn1.2}, the output at time $n$ is $\mathbf{y}(n)=CA^n\mathbf{x}(0)$.  The collected outputs $\mathbf{y}=[\mathbf{y}^\intercal(0),\cdots,\mathbf{y}^\intercal(N-1)]^\intercal$ over $N$ iterations are described by a linear transformation of the initial state
		\begin{equation}\label{OMatrix1}
			\mathbf{y}=O_{(A,C)}\mathbf{x}(0)
		\end{equation}
		where $O_{(A,C)}$ is the \definition{observability matrix}
		\begin{equation}\label{OMatrix2}
			O_{(A,C)}=\left[\begin{array}{ccc}(CA^0)^\intercal & \cdots & (CA^{N-1})^\intercal\end{array} \right]^\intercal
		\end{equation}
		and $N$ is the number of state variables.  Observability of a system can be inferred from the observability matrix as stated in Theorem~\ref{BackgroundConcepts:SystemTheory:ObservabilityThm}.
			
			%***Observability Theorem***
			\begin{theorem}[Observability \cite{SysTheo}]\label{BackgroundConcepts:SystemTheory:ObservabilityThm}
			
			The $N$ state linear, time-invariant system described by matrices $(A,C)$ is observable if and only if $\rank(O_{(A,C)})=N$.
			
			\end{theorem}
			
			In contexts involving an interval $[0,T-1]$ other than the first $N$ iterations, the following notations apply.  The collected outputs $\mathbf{y}_{[0,T-1]}=[\mathbf{y}^\intercal(0),...,\mathbf{y}^\intercal(T-1)]^\intercal$ with no input over $T$ iterations are described by a linear transformation of the initial state
		\begin{equation}\label{OMatrix1-1}
			\mathbf{y}_{[0,T-1]}=O_{(A,C),[0,T-1]}\mathbf{x}(0)
		\end{equation}
		such that
			\begin{equation}\label{OMatrix2-2}
			O_{(A,C),[0,T-1]}=\left[\begin{array}{ccc}(CA^0)^\intercal & \cdots & (CA^{T-1})^\intercal\end{array} \right]^\intercal
		\end{equation}
		with $[0,T-1]$ explicitly written for specificity.
			
			%***Controllability***
			A well known dual to the concept of observability is the concept of controllability for systems driven by inputs. Under appropriate conditions on the system matrices, these concepts are dual and results for observability hold with corresponding adaptation to controllability. This paper is primarily concerned with ensuring that the network systems designed are observable.   Duality allows the techniques developed here to apply in situations requiring the design of controllable coordinated network systems. In the subsequent text, we consider only observability and will not address controllability.

		\subsection{Structural System Theory}\label{BackgroundConcepts:StructuralSystems}
	
			%***Structural Systems***
			Because the dynamical system that describes network coordination  must respect local network connections as represented in a directed graph, analysis of how network structure affects system properties provides design insights.  Denote by $(\tilde{A},\tilde{C})$ a pair of structural matrices composed of entries that are zero or one, with $\tilde{A}\in \{0,1\}^{N\times N}$ and $\tilde{C} \in \{0,1\}^{M\times N}$.  Structural system theory examines the general system properties of all dynamic matrix pairs $(A,C)$ that respect the structure $(\tilde{A},\tilde{C})$ in the following sense.  An entry of $(A,C)$ is zero if the corresponding entry of $(\tilde{A},\tilde{C})$ is zero, while an entry of $(A,C)$ is an arbitrary parameter if the corresponding entry of $(\tilde{A},\tilde{C})$ is one.  In a straightforward way, $(\tilde{A},\tilde{C})$ may be summarized by a directed graph $\mathcal{D}(\tilde{A},\tilde{C})$.  Nodes of the graph for each of the $N$ system state variables have connections described by $\tilde{A}$, with a directed link from state $j$ to state $i$ if and only if $\tilde{A}_{ij}=1$.  Nodes of the graph for each of the $M$ system output variables have connections to states as described by $\tilde{C}$, with a directed link from state $j$ to output $i$ if and only if $\tilde{C}_{ij}=1$.
			
			%***Structural Observability***
			%***Structural Controllability***
			Specifically, conditions on the network structure that ensure recoverability of the initial state are desired.  Several results concerning structural systems are discussed in the survey paper~\cite{StructSysSurvey}, including conditions for structural observability.  The pair $(\tilde{A},\tilde{C})$ is said to be \definition{structurally observable} if there is an observable pair $(A,C)$ that respects the structure $(\tilde{A},\tilde{C})$~\cite{StructControl}.  Additionally, if $(\tilde{A},\tilde{C})$ is structurally observable, nearly all realizations that respect the structure over suitable fields, such as $\mathbb{R}$ or $\mathbb{C}$, are observable in the sense that the set of feasible unobservable realizations must have measure zero~\cite{StructControl}.
%Similarly, the pair $(\tilde{A},\tilde{B})$ is said to be \definition{structurally controllable} if there is a controllable pair $(A,B)$ that respects the structure $(\tilde{A},\tilde{B})$~\cite{StructControl}.  As with the relationship between observability and controllability, structural controllability exhibits duality with structural observability.  Hence, discussion focuses on structural observability in this paper while also applying to formulations involving structural controllability.
			
			%***Output Cacti***
			A result critical to this work, Theorem~\ref{BackgroundConcepts:StructuralSystems:StructuralObservabilityThm} shows that structural observability of $(\tilde{A},\tilde{C})$ is equivalent to the existence of an output cactus patch, a specific type of graph defined below, that spans $\mathcal{D}(\tilde{A},\tilde{C})$~\cite{StructSysSurvey}.  \definition{Output cacti} are defined recursively.  Consider a collection of nodes labeled either as state nodes or as output nodes.  An \definition{output stem} graph consists of an output node and a single directed elementary path potentially containing several state nodes rooted at the output node, with all links directed toward the output node.  All output stems are defined to be output cacti.  Furthermore, any existing output cactus to which a directed cycle of state nodes has been attached via a directed link from one node in the cycle to any node of the output cactus is also an output cactus.  Finally, a union of node-disjoint output cacti is an \definition{output cactus patch}.
			
			%***Structural Observability and Output Cacti Theorem***
			
			\begin{theorem}[Structural Observability~\cite{StructControl}\cite{StructControl2}]\label{BackgroundConcepts:StructuralSystems:StructuralObservabilityThm}
			
				The structural system matrix pair $(\tilde{A},\tilde{C})$ is structurally observable if and only if the structural system graph $\mathcal{D}(\tilde{A},\tilde{C})$ is spanned by an output cactus patch.
			
			\end{theorem}
			
			\vspace{-8pt}
	%***Network description section***
	\section{Formal Network Description}\label{NetworkDescription}
	
\begin{figure}[t]
	
		\begin{floatrow}
		
			\floatbox{figure}[\linewidth][\FBheight][t]
			{}
			{\includegraphics[width=\linewidth]{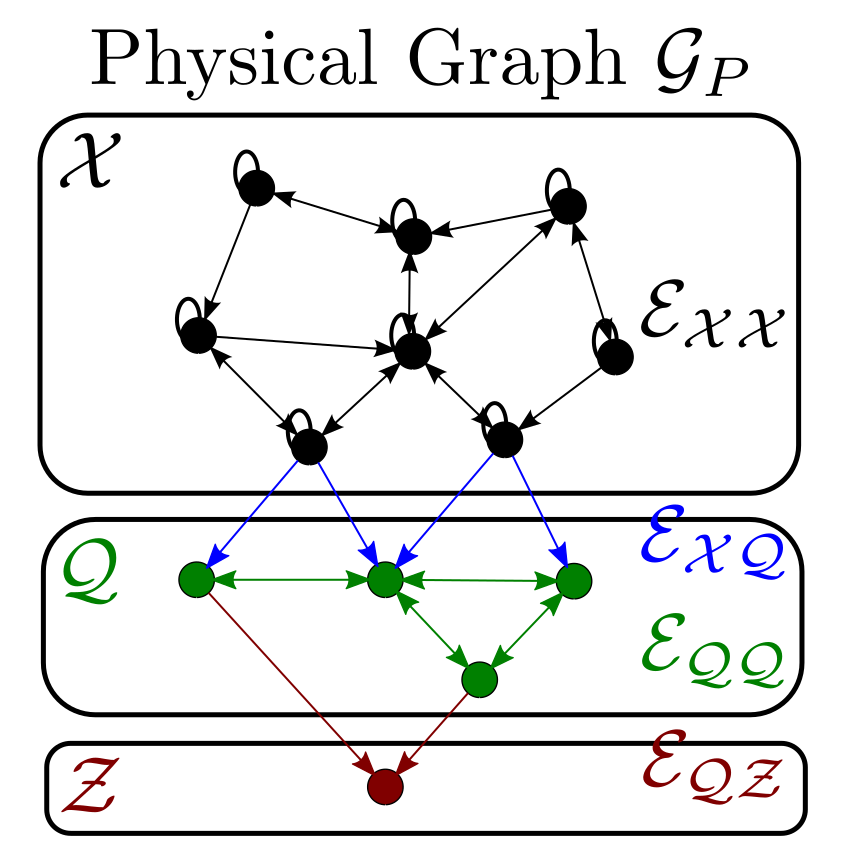}
			\caption{Physical connectivity graph $\mathcal{G}_P$ showing feasible links among sensor nodes $\mathcal{X}$ (black), backbone nodes $\mathcal{Q}$ (green), and central fusion node $\mathcal{Z}$ (red).}
			\label{NetworkExampleA}}
			
			\floatbox{figure}[\linewidth][\FBheight][t]
			{}
			{\includegraphics[width=\linewidth]{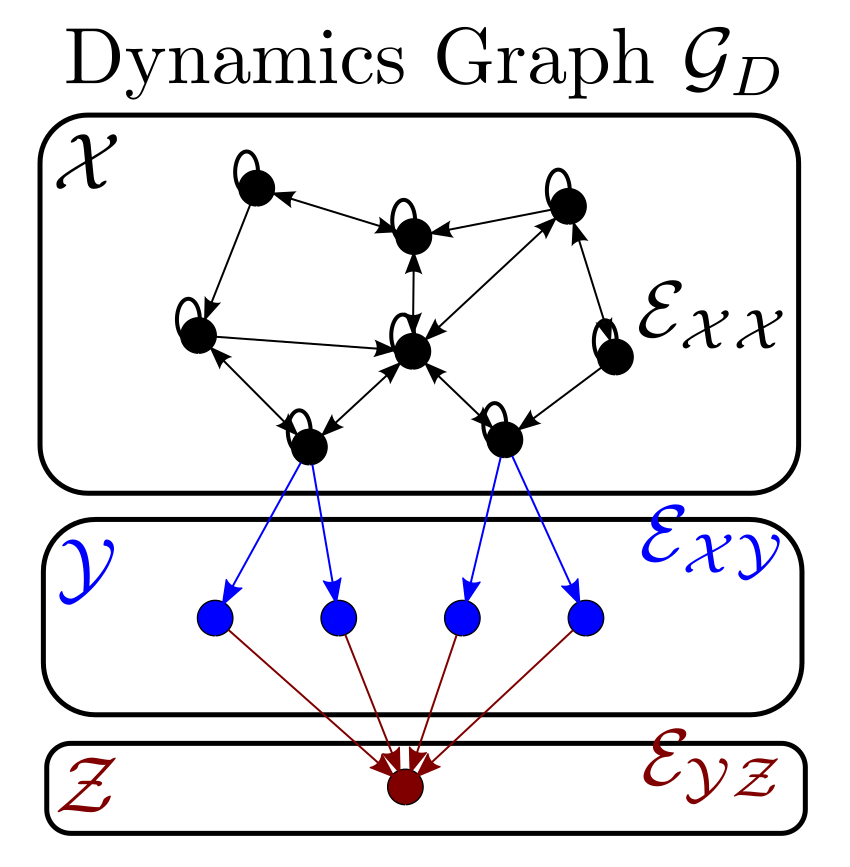}
			\caption{Dynamic system graph $\mathcal{G}_D$ showing feasible computation links among state nodes $\mathcal{X}$ (black), output nodes $\mathcal{Y}$~(blue), and central node $\mathcal{Z}$~(red).}
			%, derived from $\mathcal{G}_P$,
			\label{NetworkExampleB}}
		
		\end{floatrow}
	
	\end{figure}

		%***Node types in physical communication network***
		To formally describe the operation of the networks under consideration, we distinguish between two closely related network structures, namely, the physical network $\mathcal{G}_{P}$ describing link connections and the underlying dynamic system network $\mathcal{G}_D$ describing the relationships among sensor states and network outputs.  The physical network is composed of three types of nodes:  a set $\mathcal{X}$ of \definition{sensor nodes}, a set $\mathcal{Q}$ of \definition{backbone nodes}, and a \definition{fusion center node} $\mathcal{Z}=\{z\}$.  In this network, sensor nodes contain data and form a local state dynamics.  Backbone nodes accept outputs from the sensor nodes and route them to the fusion center node, which aggregates data for state observation tasks.   Network connectivity is described by a directed, weighted graph $\mathcal{G}_P=(\mathcal{V}_P,\mathcal{E}_P,w_P)$ where $\mathcal{V}_P=\mathcal{X}\cup\mathcal{Q}\cup\mathcal{Z}$ is the set of all nodes, $\mathcal{E}_P$ is the set of directed feasible links, and $w_P:\mathcal{E}_P\rightarrow \mathbb{R}^+$ is the weight function describing individual directed link cost.  Self-links are assumed to always be possible at zero-cost $w_P(x,x)=0$ for all $x\in \mathcal{X}$.  Figure~\ref{NetworkExampleA} provides an example of such a network, showing the sensor nodes (black), backbone nodes (green), and fusion center node (red).  Conceptually, this network divides into two subnetworks, the sensing subnetwork and the backbone subnetwork.
		
%		\begin{figure}[t]
%			\centering
%			\begin{subfigure}[t]{.46\textwidth}
%				\centering
%				\includegraphics[width=\textwidth]{NetworkExampleGP.png}
%				\caption{Example physical communication graph $\mathcal{G}_P$ showing feasible communication links among sensor nodes $\mathcal{X}$, backbone nodes $\mathcal{Q}$, and data fusion center node $\mathcal{Z}$.}
%				\label{NetworkExampleA}
%			\end{subfigure}\quad
%			\begin{subfigure}[t]{.46\textwidth}
%				\centering
%				\includegraphics[width=\textwidth]{NetworkExampleGD.png}
%				\caption{Example dynamic system graph $\mathcal{G}_D$, derived from $\mathcal{G}_P$, showing feasible computation links among state nodes $\mathcal{X}$, backbone nodes $\mathcal{Y}$, and aggregator node $\mathcal{Z}$.}
%				\label{NetworkExampleB}
%			\end{subfigure}
%			\caption{}
%		\label{NetworkExample}
%	
%	\end{figure}
		
		%***Sensor subnetwork***
		The \definition{sensor subnetwork} $\mathcal{G}_S=(\mathcal{V}_S,\mathcal{E}_S,w)$ consists of all directed links between two sensor nodes or from a sensor node to a backbone node.  Thus, $\mathcal{V}_S=\mathcal{X}\cup\mathcal{Q}$ and $\mathcal{E}_S=\mathcal{E}_{\mathcal{XX}}\cup\mathcal{E}_{\mathcal{XQ}}$ where $\mathcal{E}_{\mathcal{XX}}=\mathcal{E}_P\cap(\mathcal{X}\times\mathcal{X})$ and $\mathcal{E}_{\mathcal{XQ}}=\mathcal{E}_P\cap(\mathcal{X}\times\mathcal{Q})$.  Each sensor $x_i\in\mathcal{X}$ makes a single initial scalar measurement $\boldsymbol\alpha_i$ from field $\mathbb{F}$ and maintains a single scalar variable of state $\mathbf{x}_i(n)$ over discrete time, with initial state $\mathbf{x}_i(0)={\boldsymbol\alpha}_i$.  At every iteration, the network updates the state variable of each sensor node using local data according to a linear combination as described by a matrix $A\in \mathbb{F}^{N\times N}$ such that $\mathbf{x}(n+1)=A\mathbf{x}(n)$  where $N=|\mathcal{X}|$.  Each row of $A$ indicates which neighboring states are used to compute the updated value of a given state and the coefficients of the linear combination.  In this way, each sensor need only know the values in the corresponding row of $A$ rather than the entire network topology and dynamics.  When $A_{ij}\neq 0$, sensor node $x_j$ is linked to $x_i$, incurring cost.  Otherwise, no link exists between the two nodes.
		
		%***Backbone subnetwork***
		The \definition{backbone subnetwork} $\mathcal{G}_B=(\mathcal{V}_B,\mathcal{E}_B,w)$ consists of all directed links between two backbone nodes or from a backbone node to the fusion center.  Thus, $\mathcal{V}_B=\mathcal{Q}\cup\mathcal{Z}$ and $\mathcal{E}_B=\mathcal{E}_{\mathcal{QQ}}\cup\mathcal{E}_{\mathcal{QZ}}$ where $\mathcal{E}_{\mathcal{QQ}}=\mathcal{E}_P\cap(\mathcal{Q}\times \mathcal{Q})$ and $\mathcal{E}_{\mathcal{QZ}}=\mathcal{E}_P\cap(\mathcal{Q}\times \mathcal{Z})$.  Some of the states are output (sensed) through the network backbone nodes in $\mathcal{Q}$ and available at the central node $z$ as the vector $\mathbf{y}(n)$.  It is assumed that the backbone subnetwork operates much faster than the sensor subnetwork iterations, such that $\mathbf{y}(n)$ is available to $z$ without delay.  For link cost efficiency, the backbone network uses the least costly path to the central node.  These outputs can be described by a matrix $C\in\mathbb{F}^{M\times N}$ such that $\mathbf{y}(n)=C\mathbf{x}(n)$ where $M=|\mathcal{E}_{\mathcal{XQ}}|$ is the number of feasible connections between the sensors and backbone nodes.  Each row $i$ of $C$ either is composed of all zeros, indicating an output is not made over the corresponding connection and incurring no cost, or has a single nonzero entry in column $j$, indicating state $x_j$ is output over the connection and incurring cost.
		
		%***Node types in dynamic system network***
		Consider the system dynamics network $\mathcal{G}_D$ that describes the relationships among sensor states, backbone outputs, and fusion center in the physical network.  The sensor nodes $\mathcal{X}$ of the physical network correspond to the \definition{state nodes}, and maintain the same links $\mathcal{E}_{\mathcal{XX}}$ with $w_D(x_1,x_2)=w_P(x_1,x_2)$ for $x_1,x_2\in\mathcal{X}$.  Additionally, the fusion center node $\mathcal{Z}=\{z\}$ appears in this network.  Because the backbone nodes may output the states from multiple sensor nodes, backbone nodes do not correspond to only one output.  Rather, each link in $\mathcal{E}_{\mathcal{XQ}}$ from a sensor node to a backbone node represents a potential output of the system.  Thus, let a set $\mathcal{Y}$ of \textbf{output nodes} be indexed by $\mathcal{E}_{\mathcal{XQ}}$ such that $y_{(x,q)}\in\mathcal{Y}$ corresponds to $(x,q)\in \mathcal{E}_{\mathcal{XQ}}$.  Each output node $y_{(x,q)}$ will be linked to the corresponding sensor node $x$ with weight $w_D(x,y_{(x,q)})=w_P(x,q)$ and to the central node with weight $w_D(y_{(x,q)},z)=w_{\spath}(q,z;\mathcal{E}_{B})$ where $w_{\spath}(q,z;\mathcal{E}_{B})$ is the weight of the shortest path from $q$ to $z$ over the backbone links $\mathcal{E}_B$.  Hence, $\mathcal{E}_{\mathcal{XY}}=\{(x,y_{(x,q)})|(x,q)\in\mathcal{E}_{\mathcal{XQ}}\}$ and $\mathcal{E}_{\mathcal{YZ}}=\{(y_{(x,q)},z)|(x,q)\in\mathcal{E}_{\mathcal{XQ}}\}$.  Note that state nodes may connect to multiple output nodes, but each output node corresponds to only one state node.  Also, note that output nodes do not have links to each other, so the output nodes together with the fusion center form a star topology subnetwork.  Thus, the graph that describes the state and output update dependencies is $\mathcal{G}_D=(\mathcal{X}\cup\mathcal{Y}\cup\mathcal{Z},\mathcal{E}_{\mathcal{XX}} \cup\mathcal{E}_{\mathcal{XY}} \cup\mathcal{E}_{\mathcal{YZ}},w_D)$.  As with the physical network, a link only contributes cost to the network operation if it is required for updating states and outputs.  The nodes, links, and weights of this network are entirely derived from the physical network $\mathcal{G}_P$, and Figure~\ref{NetworkExampleB} provides an example of such a modification.  The figure shows the organization into state nodes (black), output nodes (blue), and the central node (red).
		
		%***Design constraints (observability)***
%		The network state $\mathbf{x}(n)$ and the network output $\mathbf{y}(n)$ behave according to the dynamical system
%	\begin{alignat}{4}
%		&\mathbf{x}({n+1})&&=A\mathbf{x}(n)\label{dyn1} \\
%		&\mathbf{y}({n}) &&=C\mathbf{x}(n)\label{dyn2}
%	\end{alignat}
%	where $\mathbf{x}(0)={\boldsymbol\alpha}$ is the vector of sensor measurements.  The network seeks to communicate all sensor measurements ${\boldsymbol\alpha}$ to the central node by reconstruction of the initial system state from the network backbone outputs forwarded to the central node.  Therefore, observability of the system must be ensured when designing the matrix pair $(A,C)$, which translates to the full rank condition on $O_{(A,C)}$.  Furthermore, the fusion center must be aware of the entire network dynamics so that the observability matrix is known and that estimation of the initial state may be performed.

The network state $\mathbf{x}(n)$ and the network output $\mathbf{y}(n)$ behave according to the dynamical system~\eqref{dyn1.1}-\eqref{dyn1.2} with
%	\begin{alignat}{4}
%		&\mathbf{x}({n+1})&&=A\mathbf{x}(n)\label{dyn1} \\
%		&\mathbf{y}({n}) &&=C\mathbf{x}(n)\label{dyn2}
%	\end{alignat}
	$\mathbf{x}(0)={\boldsymbol\alpha}$ being the initial state.  Observability of the system must be ensured when designing the matrix pair $(A,C)$ in~\eqref{dyn1.1}-\eqref{dyn1.2}, which translates to the full rank condition on $O_{(A,C)}$.  Furthermore, the fusion center must be aware of the entire network dynamics so that the observability matrix is known and state observation tasks may be performed.
	
		%***Design constraints (link feasibility)***		
		Additionally, each node must only access the state from its neighbors in $\mathcal{G}_P$ to update its state or output values but does not necessarily use information from every neighbor.  Let $\mathcal{G}_P(A,C)=(\mathcal{V}_P,\mathcal{E}_P(A,C),w_P)$ be the subgraph of $\mathcal{G}_P$ defined by the set $\mathcal{E}_P(A,C)$ of physical links required for the computation of \eqref{dyn1.1}-\eqref{dyn1.2} and outputs along the least costly backbone path.  Note that $\mathcal{E}_P(A,C)=\mathcal{E}_{\mathcal{XX}}(A)\cup \mathcal{E}_{\mathcal{XQ}}(C) \cup \mathcal{E}_{B}(C)$ with links in $\mathcal{E}_{\mathcal{XX}}(A)$ used to compute~\eqref{dyn1.1},  links in $\mathcal{E}_{\mathcal{XQ}}(C)$ used in~\eqref{dyn1.2}, and  links in $\mathcal{E}_{B}(C)$ used in the shortest path from every used backbone node $q$ directly connected to some sensor $x$ by $(x,q)\in\mathcal{E}_{\mathcal{XQ}}(C)$.  For feasibility of network operations, the subset constraints $\mathcal{E}_{\mathcal{XX}}(A)\subseteq \mathcal{E}_{\mathcal{XX}}$, $\mathcal{E}_{\mathcal{XQ}}(C)\subseteq \mathcal{E}_{\mathcal{XQ}}$, and $\mathcal{E}_{B}(C)\subseteq \mathcal{E}_B$ must hold.
		
		%***Design objective and problem contrasts***
		Hence, these link feasibility and system observability constraints must be applied when designing the network dynamics $(A,C)$.  Furthermore, minimization of the sum total cost
	\begin{equation}\label{objective}
		F(A,C)=\sum_{\begin{array}{c}e\in \mathcal{E}_P(A,C)\end{array}}{ w_P(e)}
	\end{equation}
	of all used network physical links is desired, including sensor state updates, sensor output production, and backbone activity.  Section~\ref{DesignProblem} formulates a robust version of the minimum cost network design problem with this objective function and set of constraints.
		
	%***Network design problem formulation and main results***
	\section{Network Design Problem}\label{DesignProblem}

		%***Preliminary introduction***
		The problem examined in this paper concerns optimal design of networks operating according to the description in Section~\ref{NetworkDescription}.  As equation \eqref{objective} shows, the underlying system dynamics determines the physical link cost of operating such networks.  A less complete formulation of this problem appeared in \cite{PequitoKruzickEUSIPCO2013}, which provides an efficient solution algorithm based on minimum spanning tree methods for the simpler and much more fragile case in which only one backbone node exists and no node failures may occur.  In contrast, this paper significantly generalizes this previous work by formulating the optimal design problem in the context of nontrivial backbone subnetworks and sensor node failures.
		
		%***Unified formulation***
		Therefore, this section examines minimization of the objective function in \eqref{objective} with respect to the dynamic matrices $(A,C)$ subject to the aforementioned system observability and link-connectivity constraints under added robustness requirements arising from sensor failures.  Specifically, consider optimal design of a link feasible network where system observability must be guaranteed,  while any subset of sensor nodes $\mathcal{U}\subset \mathcal{X}$ of size $|\mathcal{U}|\leq k$ less than the robustness design parameter $k$ experiences total failure before coordination begins.  In this way, the original states of nodes that did not fail could still be recovered from the collected outputs.  Sensor node failure results in the elimination of associated rows and columns in the network dynamics matrices.  Note that the row of $A$, columns of $A$, and columns of $C$ are indexed by the sensor nodes $\mathcal{X}$, while the rows of $C$ are indexed by $\mathcal{E}_{\mathcal{XQ}}$. Denote by ${A}_{[\mathcal{X}\backslash\mathcal{U},\mathcal{X}\backslash\mathcal{U}]}$ and ${C}_{[\mathcal{E}_{\mathcal{XQ}},\mathcal{X}\backslash\mathcal{U}]}$ submatrices of $A$ and $C$, respectively, in which rows and columns corresponding to $\mathcal{U}$ have been removed.  This yields the optimization problem appearing below.
		%***First formulation***
		\begin{alignat}{4}
			&\argmin_{{A},{C}} \quad &&	F({A},{C}) \label{DesignProblem:V1:Eq1} \\
			&\st \quad &&\mathcal{E}_{\mathcal{XX}}({A})\quad\subseteq \quad \mathcal{E}_{\mathcal{XX}} \label{DesignProblem:V1:Eq2} \\
			& \quad &&\mathcal{E}_{\mathcal{XQ}}({C})\quad\subseteq \quad \mathcal{E}_{\mathcal{XQ}} \label{DesignProblem:V1:Eq3}\\
			& \quad &&\mathcal{E}_{B\hphantom{B}}({C})\quad\subseteq \quad \mathcal{E}_{B} \label{DesignProblem:V1:Eq4}\\
			& \quad &&({A}_{[\mathcal{X}\backslash\mathcal{U},\mathcal{X}\backslash\mathcal{U}]},{C}_{[\mathcal{E}_{\mathcal{XQ}},\mathcal{X}\backslash\mathcal{U}]}) \label{DesignProblem:V1:Eq5}\\
			& \quad &&\mathrlap{\textrm{observable}}\hphantom{\textrm{spanned by output cactus patch}} \nonumber\\
			& \quad &&\textrm{for all }\mathcal{U}\subset \mathcal{X}\textrm{ with }|\mathcal{U}|\leq k\nonumber
		\end{alignat}
		
		%***Why this problem could be difficult***
		The optimization described by \eqref{DesignProblem:V1:Eq1}-\eqref{DesignProblem:V1:Eq5} appears seemingly difficult due to the large number of observability constraints that must be satisfied.  Recall that Theorem~\ref{BackgroundConcepts:SystemTheory:ObservabilityThm} states the equivalence between observability of $(A,C)$ and the condition $\rank(O_{(A,C)})=N$.  Thus, the constraint given in \eqref{DesignProblem:V1:Eq5} translates to $\rank(O_{({A}_{[\mathcal{X}\backslash\mathcal{U},\mathcal{X}\backslash\mathcal{U}]}, {C}_{[\mathcal{E}_{\mathcal{XQ}},\mathcal{X}\backslash\mathcal{U}]}})=N-|\mathcal{U}|$ for all $\mathcal{U}\subset \mathcal{X}$ with $|\mathcal{U}|\leq k$.  Because there are potentially very many ways to form subsets of $\mathcal{X}$ of size at most $k$, a large number of rank constraints must be satisfied.  As such, it is not immediately clear that this problem is efficiently solvable.  Furthermore, it is not obvious how to check that a solution exists for a given $k$ or to determine the largest $k$ for which a solution exists.  Later in the paper, after additional discussion of this problem, Remark \ref{DesignProblem:ExistenceRmk} addresses existence of the solution in terms of flow problems.  However, the optimization problem must first be related more closely to the combinatorial structure of the graph.
		
		%***Decouple Problems***
		A structural systems approach renders the problem tractable by decoupling the problem of finding the optimal structure for the dynamic matrices from the problem of finding an observable instantiation of the optimal structure.  Note that the objective function in \eqref{DesignProblem:V1:Eq1} only depends on which links were used, as do the link feasibility constraints \eqref{DesignProblem:V1:Eq2}-\eqref{DesignProblem:V1:Eq5}.  Thus, it depends on only the zero-nonzero structure $(\tilde{A},\tilde{C})$ of $(A,C)$ and can be restated in structural terms.  By the definition of structural observability, if $(\tilde{A},\tilde{C})$ is structurally observable then an observable instantiation $(A,C)$ respecting that structure must exist and may be found subsequently.  Therefore, the final constraint can be rewritten to guarantee structural observability under limited node failures.  The problem appears below with this reformulation.
		%***Second Formulation***
		\begin{alignat}{4}		
			&\argmin_{\tilde{A},\tilde{C}} \quad &&	F(\tilde{A},\tilde{C}) \label{DesignProblem:V2:Eq1} \\
			&\st \quad &&\mathcal{E}_{\mathcal{XX}}(\tilde{A})\quad\subseteq \quad \mathcal{E}_{\mathcal{XX}} \label{DesignProblem:V2:Eq2} \\
			& \quad &&\mathcal{E}_{\mathcal{XQ}}(\tilde{C})\quad\subseteq \quad \mathcal{E}_{\mathcal{XQ}} \label{DesignProblem:V2:Eq3}\\
			& \quad &&\mathcal{E}_{B\hphantom{B}}(\tilde{C})\quad\subseteq \quad \mathcal{E}_{B} \label{DesignProblem:V2:Eq4}\\
			& \quad &&({\tilde{A}}_{[\mathcal{X}\backslash\mathcal{U},\mathcal{X}\backslash\mathcal{U}]},{\tilde{C}}_{[\mathcal{E}_{\mathcal{XQ}},\mathcal{X}\backslash\mathcal{U}]}) \label{DesignProblem:V2:Eq5}\\
			& \quad &&\mathrlap{\textrm{structurally observable}}\hphantom{\textrm{spanned by output cactus patch}} \nonumber\\
			& \quad &&\textrm{for all }\mathcal{U}\subset \mathcal{X}\textrm{ with }|\mathcal{U}|\leq k\nonumber
		\end{alignat}
		
		%***Reduction to spanning cactus***
		Replacement of the observability constraint with a structural observability constraint may not appear, at first, to suggest a solution.  However, by appealing to Theorem~\ref{BackgroundConcepts:StructuralSystems:StructuralObservabilityThm}, the structural observability constraint can be replaced by the equivalent condition that the associated directed graph be spanned by an output cactus patch after any set of at most $k$ sensor node deletions are applied.  Hence, the original analytic rank constraint has been transformed to a combinatorial structural constraint in the problem below.
		%***Third formulation***
		\begin{alignat}{4}
			&\argmin_{\tilde{A},\tilde{C}} \quad &&	F(\tilde{A},\tilde{C}) \label{DesignProblem:V3:Eq1} \\
			&\st \quad &&\mathcal{E}_{\mathcal{XX}}(\tilde{A})\quad\subseteq \quad \mathcal{E}_{\mathcal{XX}} \label{DesignProblem:V3:Eq2} \\
			& \quad &&\mathcal{E}_{\mathcal{XQ}}(\tilde{C})\quad\subseteq \quad \mathcal{E}_{\mathcal{XQ}} \label{DesignProblem:V3:Eq3}\\
			& \quad &&\mathcal{E}_{B\hphantom{B}}(\tilde{C})\quad\subseteq \quad \mathcal{E}_{B} \label{DesignProblem:V3:Eq4}\\
			& \quad &&D({\tilde{A}}_{[\mathcal{X}\backslash\mathcal{U},\mathcal{X}\backslash\mathcal{U}]},{\tilde{C}}_{[\mathcal{E}_{\mathcal{XQ}},\mathcal{X}\backslash\mathcal{U}]}) \label{DesignProblem:V3:Eq5}\\
			& \quad &&\textrm{spanned by output cactus patch} \nonumber\\
			& \quad &&\textrm{for all }\mathcal{U}\subset \mathcal{X}\textrm{ with }|\mathcal{U}|\leq k \nonumber
		\end{alignat}
		
		%***Algorithm presentation***
			\begin{figure*}[t]\TopFloatBoxes
		\begin{floatrow}
			
			\floatbox{table}[\linewidth][\FBheight][t]
			{}
			{
				\begin{tabular}{@{} p{\widthof{Output:~~}+5pt} @{} p{\linewidth-\widthof{Output:~~}-5pt} @{}}
					\toprule
					\multicolumn{2}{c}{\textbf{Algorithm 1}} \\ \midrule
					
					\textbf{Input~~:} &
						\parbox[t]{\linewidth}{Physical network communication cost graph\\
						$\mathcal{G}_P=(\mathcal{X}\cup\mathcal{Q}\cup\mathcal{Z}, \mathcal{E}_S\cup \mathcal{E}_B, w_P)$ with\\
						$\mathcal{E}_S=\mathcal{E}_{\mathcal{XX}}\cup\mathcal{E}_{\mathcal{XQ}}$ and $\mathcal{E}_{B}=\mathcal{E}_{\mathcal{QQ}}\cup\mathcal{E}_{\mathcal{QZ}}$\\
						Required degree of robustness $k$} \\
				
				\textbf{Step ~1:} &
					\parbox[t]{\linewidth}{Compute the least costly path $\spath(q,z;\mathcal{E}_B)$ from each $q\in \mathcal{Q}$ to the central node $z$ over the backbone subnetwork links $\mathcal{E}_B$.  This may be accomplished by computing the shortest path spanning arborescence of $\mathcal{G}_{B}=(\mathcal{Q}\cup\mathcal{Z},\mathcal{E}_B,w_P)$ rooted at $z$ \cite{CombinatorialOpt}.  Let the weight of the shortest path from $q$ to $z$ be $w_{\spath}(q,z;\mathcal{E}_B)$.} \\
				
				\textbf{Step ~2:} &
					\parbox[t]{\linewidth}{Generate the dynamic system cost graph $\mathcal{G}_D=(\mathcal{X}\cup\mathcal{Y}\cup\mathcal{Z},\mathcal{E}_{\mathcal{XX}} \cup \mathcal{E}_{\mathcal{XY}} \cup \mathcal{E}_{\mathcal{YZ}},w_D)$ from the physical links cost graph $\mathcal{G}_P$ by removing the backbone nodes $\mathcal{Q}$ and adding the output nodes $\mathcal{Y}$ as described in Section~\ref{NetworkDescription}.
					For each $(x_1,x_2)\in\mathcal{E}_{\mathcal{XX}}$, $w_D(x_1,x_2)=w_P(x_1,x_2)$.\\
					The link $(x,y_{(x,q)})\in\mathcal{E}_{\mathcal{XY}}$ if and only if $(x,q)\in\mathcal{E}_{\mathcal{XQ}}$, with $w_D(x,y_{(x,q)})=w_P(x,q)$.  For each $y_{(x,q)}\in \mathcal{Y}$ and $\mathcal{Z}=\{z\}$, $(y_{(x,q)},z)\in\mathcal{E}_{YZ}$ and $w_D(y_{(x,q)},z)=w_{\spath}(q,z;\mathcal{E}_B)$.} \\
				
				\textbf{Step ~3:} &
					\parbox[t]{\linewidth}{Modify $\mathcal{G}_D$ to produce the graph $\mathcal{G}_D'=(\mathcal{X}\cup\mathcal{Y}\cup\mathcal{Z},\mathcal{E}_{\mathcal{XX}} \cup \mathcal{E}_{\mathcal{XY}} \cup \mathcal{E}_{\mathcal{YZ}},w_D')$.  For each $(x_1,x_2)\in\mathcal{E}_{\mathcal{XX}}$, $w_D'(x_1,x_2)=w_P(x_1,x_2)$.  For each $(x,y_{(x,q)})\in \mathcal{E}_{\mathcal{XY}}$ and $\mathcal{Z}=\{z\}$, $w_D'(x,y_{(x,q)})=w_P(x,q)+w_{\spath}(q,z;\mathcal{E}_B)$ and $w_D'(y_{(x,q)},z)=0$.  Note that all augmenting links of $\mathcal{G}_D'$ have tail in the sensor nodes $\mathcal{X}$.} \\
				
				\textbf{Step ~4:} &
					\parbox[t]{\linewidth}{Find the minimum $z$-rooted $(k+1)$-node-connected Steiner subgraph of $\mathcal{G}_D'$ with terminal nodes $\mathcal{X}$  using a suitable algorithm, such as that provided in \cite{FrankRootedKConnections}.  The modification suggested in Remark~\ref{DesignProblem:ComplexityRmk} may also be applied.  Refer to this graph as $\mathcal{T}^*=(\mathcal{X}\cup\mathcal{Y}\cup\mathcal{Z},\mathcal{E}_{\mathcal{T}^*},w_D')$.} \\
				
				\textbf{Step ~5:} &
					\parbox[t]{\linewidth}{Construct a graph $\mathcal{P}^*=(\mathcal{X}\cup\mathcal{Y},\mathcal{E}_{\mathcal{P}^*},w_D')$ by adding a zero-cost self-loop to each sensor node in the graph $\mathcal{T}^*$ and by removing $z$.  Hence, $\mathcal{E}_{\mathcal{P}^*}=(\mathcal{E}_{\mathcal{T}^*}\backslash\mathcal{E}_\mathcal{YZ})\cup\{(x,x)|x\in \mathcal{X}\}$.} \\
				
				\textbf{Step ~6:} &
					\parbox[t]{\linewidth}{Form $(\tilde{A}^*,\tilde{C}^*)$ such that $\mathcal{P}^*\simeq \mathcal{D}(\tilde{A}^*,\tilde{C}^*)$.\\
					Set $\tilde{A}^*_{ij}=1$ if and only if $(x_j,x_i)\in\mathcal{E}_{\mathcal{P}^*}$.\\
					Set $\tilde{C}^*_{kj}=1$ if and only if $(x_j,q_k)\in\mathcal{E}_{\mathcal{P}^*}$.} \\
				
				\textbf{Output:} &
					\parbox[t]{\linewidth}{Optimal dynamics structure $(\tilde{A}^*,\tilde{C}^*)$\\
					Corresponding physical graph $\mathcal{G}_P(\tilde{A}^*,\tilde{C}^*)$} \\
				
				\bottomrule
				
				\end{tabular}
			}
			\floatbox{figure}[\linewidth][\FBheight][t]
			{}
			{\centering
			\begin{tabular}{@{} c @{}}
			%\toprule
			\vspace{\abovetopsep}
			\vspace{\heavyrulewidth}
			\vspace{\belowrulesep}
			\phantom{Algorithm 1} \\
			\vspace{\aboverulesep}
			\vspace{\lightrulewidth}
			\vspace{\belowrulesep}
			\includegraphics[width=.96\linewidth]{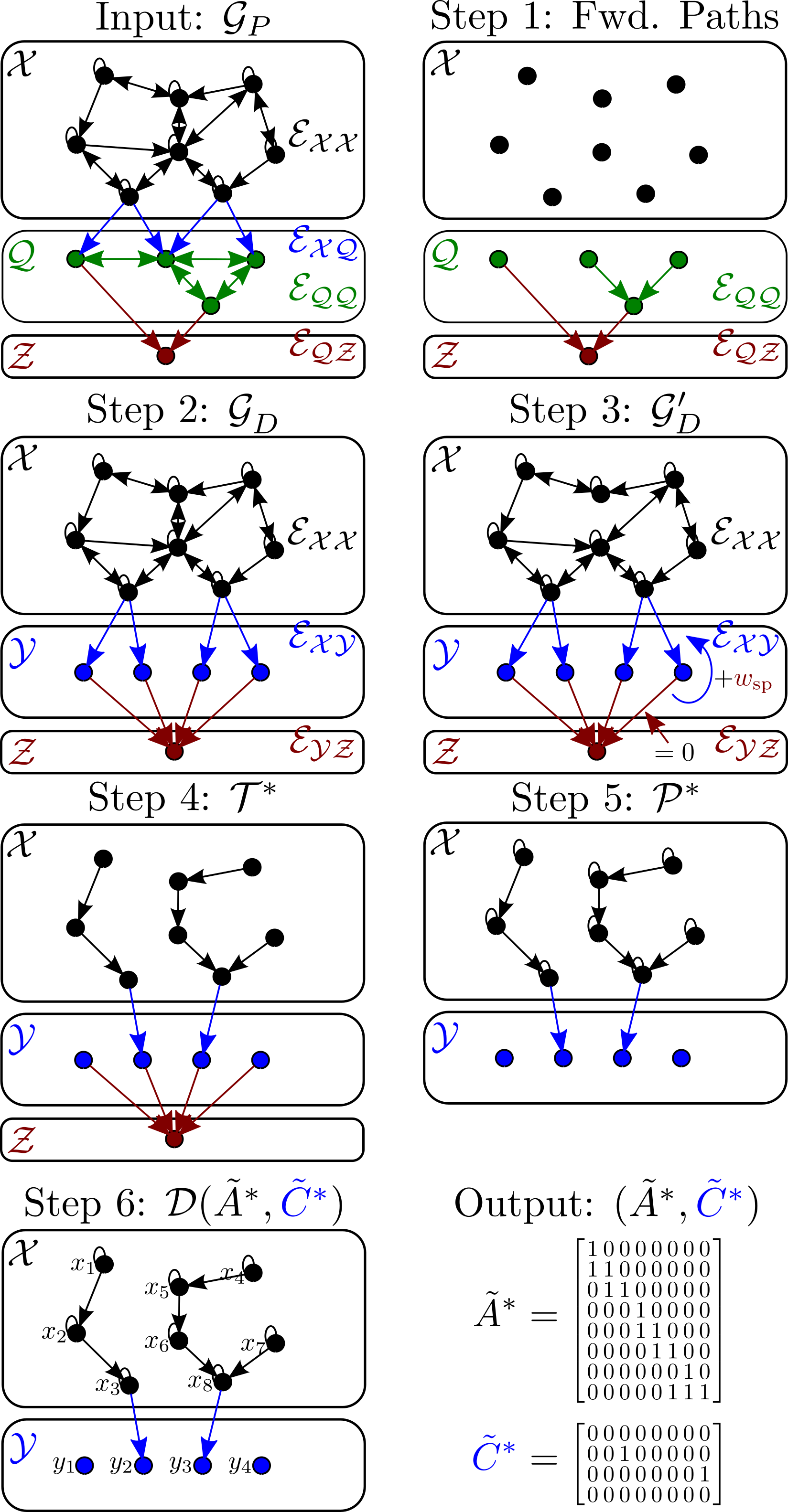}
			\end{tabular}
			\caption{The illustration demonstrates the steps of Algorithm~1 for an example input graph $\mathcal{G}_P$ and $k=0$, resulting in a $\mathcal{Y}$-rooted branching with added loops.}
			\label{Alg1Fig}
			}
		\end{floatrow}
	\end{figure*}
	
			%***Linking paragraph***
		This optimization problem is solved by Algorithm~1, which takes as input the graph $\mathcal{G}_P$ of all possible physical network links.  Through six steps, illustrated in Figure \ref{Alg1Fig}, it produces as output an optimal dynamics structure $(\tilde{A}^*,\tilde{C}^*)$ for the operation of the network.  This also defines the subgraph $\mathcal{G}_P(\tilde{A}^*,\tilde{C}^*)$ of $\mathcal{G}_P$ that is necessary for the operation of the network, assuming backbone forwarding occurs along the least costly path.  Practical discussion of how to find an observable instantiation of this structure appears later in this section.
		
		%***Algorithm discussion**
		Algorithm~1 begins by first finding the cost of the fusion center accessing a single output in the physical links cost graph $\mathcal{G}_P$ from each backbone node by computing the shortest path spanning arborescence of the backbone subnetwork in Step~1.  Subsequently, Step~2 generates the dynamic system computation cost graph $\mathcal{G}_D$, which it accomplishes by removing the backbone nodes and creating an output node $y_{(x,q)}$ for every $(x,q)\in \mathcal{E}_{\mathcal{XQ}}$.  This output node connects to only one sensor node $x$ with cost $w_D(x,y_{(x,q)})=w_P(x,q)$ and the fusion center $z$ with cost $w_D(y_{(x,q)},z)= w_{\spath}(q,z;\mathcal{E}_B)$ obtained from the shortest path spanning arborescence found in Step~1.  In order to frame the optimization problem as an efficiently solvable (by submodular flow) case of the Steiner subgraph problem with $\mathcal{X}$ as the terminal node set, all augmenting links must have tail in the terminal node set.  Therefore, Step~3 produces the modified graph $\mathcal{G}_D'$ in which $w_D'(x,y_{(x,q)})=w_P(x,q)+w_{\spath}(q,z;\mathcal{E}_B)$ and $w_D'(y_{(x,q)},z)=0$.  Note that any minimal $z$-rooted Steiner subgraph of $\mathcal{G}_D'$ with terminal nodes $\mathcal{X}$ either include both $(x,y_{(x,q)})$ and $(y_{(x,q)},z)$ or includes neither $(x,y_{(x,q)})$ nor $(y_{(x,q)},z)$.  Hence, this modification does not affect the total cost of any minimal solution.  Step~4 performs the optimization step, which is a minimum $z$-rooted $(k+1)$-node-connected Steiner subgraph computation for $\mathcal{G}_D'$ with terminal nodes $\mathcal{X}$.  Because all augmenting links have tail in $\mathcal{X}$ due to the modification in Step~3, the problem is solvable in polynomial time using submodular flows \cite{FrankRootedKConnections}.  In fact, it can be solved through a maximum weighted matroid intersection algorithm \cite{FrankRootedKConnections} by the method in Remark \ref{DesignProblem:ComplexityRmk}.  Step~5 constructs a graph that is guaranteed to be spanned by an output cactus patch through the addition of zero-cost self loops to the sensor nodes.  Finally, Step~6 interprets the graph from Step~5 as a structural system, allowing an optimal structurally observable network dynamics matrix pair to be output from the algorithm.  Theorem~\ref{DesignProblem:MainThm} and the corresponding proof show that the supplied algorithm solves the minimum cost design problem for any input physical links cost network $\mathcal{G}_P$ and robustness requirement $k$ such that a solution exists.

		%***Theorem and proof***
		
		\begin{theorem}[Main Result]\label{DesignProblem:MainThm}
		
			For the  physical links cost network $\mathcal{G}_P$ and robustness requirement $k$ the structural dynamics pair $(\tilde{A}^*,\tilde{C}^*)$ output by Algorithm~1 is a solution to the optimization problem in \eqref{DesignProblem:V2:Eq1}-\eqref{DesignProblem:V2:Eq5}, provided that for each $x\in \mathcal{X}$ there is a set of at least $k+1$ internally node-disjoint directed paths in $\mathcal{G}_P$ each beginning at $x$ and ending in $Q$ and that there is a directed path from each $q\in Q$ to $z$.
		
		\end{theorem}
		
		\vspace{-10pt}
		\begin{proof}\label{DesignProblem:MainPrf}
		
			We first demonstrate the feasibility and existence of the solution $(\tilde{A}^*,\tilde{C}^*)$.  Subsequently, we show by contradiction that no other feasible solution incurring lesser cost exists.  Hence, we conclude the solution found is optimal.
			
			Note that the shortest path spanning arborescence of the backbone subnetwork computed in Step~1 must exist because there is a directed path from each $q\in \mathcal{Q}$ to $z$.  There are $k+1$ internally node-disjoint directed paths in $\mathcal{G}_P$ beginning at $x$ and ending in $\mathcal{Q}$, each of which contains a distinct link $(x_i,q_i)\in \mathcal{E}_{\mathcal{XQ}}$.  Thus, the graph $\mathcal{G}_D$ constructed in Step~2 has $k+1$ internally node-disjoint directed paths beginning at $x$ and ending in a distinct $y_{i}=y_{(x_i,q_i)}\in\mathcal{Y}$, which can be constructed from the paths in $\mathcal{G}_P$ by substituting $(x_i,y_{i})$ for the final link.  Furthermore, because each $y_i\in\mathcal{Y}$ connects directly to $z$, this implies that $\kappa_{\mathcal{G}_D}(x,z)\geq k+1$ for all $x\in\mathcal{X}$.  Since construction of $\mathcal{G}_D'$ in Step~3 does not alter connectivity, it also follows that $\kappa_{\mathcal{G}_D'}(x,z)\geq k+1$ for all $x\in\mathcal{X}$.  Hence, the minimum $z$-rooted $(k+1)$-node-connected Steiner subgraph of $\mathcal{G}_D'$ with terminal nodes $\mathcal{X}$ exists and $\mathcal{T}^*$ can be found in Step~4.  With deletion of any failing node set $\mathcal{U}\subset \mathcal{X}$ with $|\mathcal{U}|\leq k$, there is at least one path from each node in $\mathcal{X}\backslash\mathcal{U}$ to $z$ by $z$-rooted $(k+1)$-node-connectedness.  These paths form a spanning tree $\mathcal{T}_{\mathcal{U}}^*$ for the subgraph of $\mathcal{T}^*$ in which $\mathcal{U}$ has been excluded.  Because zero-cost self-loops are always permitted at every $x\in \mathcal{X}$, $\mathcal{P}^*$ may be formed in Step~5, and the subgraph of $\mathcal{P}^*$ that excludes $\mathcal{U}$ is spanned by $\mathcal{T}_{\mathcal{U}}^*$ with $z$ removed.  Let $\mathcal{P}_{\mathcal{U}}^*$ be the graph formed from $\mathcal{T}_{\mathcal{U}}^*$ by addition of zero cost self-loops to all $x\in \mathcal{X}\backslash\mathcal{U}$ and with $z$ removed.  Note that $\mathcal{P}_{\mathcal{U}}^*$ satisfies the recursive definition of an output cactus patch, with output nodes $\mathcal{Y}$, and that $\mathcal{P}_{\mathcal{U}}^*$ spans the subgraph of $\mathcal{P}^*$ with $\mathcal{U}$ excluded.  By the construction in Step~6, $\mathcal{D}(\tilde{A}^*,\tilde{C}^*)$ is isomorphic to $\mathcal{P}^*$, and, consequently, $D({\tilde{A}^*}_{[\mathcal{X}\backslash\mathcal{U},\mathcal{X}\backslash\mathcal{U}]}, {\tilde{C}^*}_{[\mathcal{E}_{\mathcal{XQ}},\mathcal{X}\backslash\mathcal{U}]})$ is spanned by an output cactus patch isomorphic to $\mathcal{P}_{\mathcal{U}}^*$ (by the same isomorphism).  Hence, by Theorem~\ref{BackgroundConcepts:StructuralSystems:StructuralObservabilityThm}, $({\tilde{A}^*}_{[\mathcal{X}\backslash\mathcal{U},\mathcal{X}\backslash\mathcal{U}]}, {\tilde{C}^*}_{[\mathcal{E}_{\mathcal{XQ}},\mathcal{X}\backslash\mathcal{U}]})$ is structurally observable for all $\mathcal{U}\subset \mathcal{X}$ with $|\mathcal{U}|<k$.  The other constraints, $\mathcal{E}_{\mathcal{XX}}(\tilde{A}^*)\subseteq \mathcal{E}_{\mathcal{XX}}$, $\mathcal{E}_{\mathcal{XQ}}(\tilde{C}^*)\subseteq \mathcal{E}_{\mathcal{XQ}}$, and $\mathcal{E}_{B}(\tilde{C}^*)\subseteq \mathcal{E}_{B}$, are also satisfied through subgraph constructions.  Therefore, $(\tilde{A}^*,\tilde{C}^*)$ is a feasible solution.
			
			Assume by way of contradiction that a feasible solution $(\tilde{A}^\dag,\tilde{C}^\dag)$ of lesser cost than $(\tilde{A}^*,\tilde{C}^*)$ with respect to the objective function $F$ exists.  That is,
			\begin{equation}\label{DesignProblem:MainPrf:Eq1}
				F(\tilde{A}^\dag,\tilde{C}^\dag)<F(\tilde{A}^*,\tilde{C}^*).
			\end{equation}
			Because $({\tilde{A}^\dag}_{[\mathcal{X}\backslash\mathcal{U},\mathcal{X}\backslash\mathcal{U}]}, {\tilde{C}^\dag}_{[\mathcal{E}_{\mathcal{XQ}},\mathcal{X}\backslash\mathcal{U}]})$ must be structurally observable for all $\mathcal{U}\subset\mathcal{X}$ with $|\mathcal{U}|\leq k$, $D({\tilde{A}^\dag}_{[\mathcal{X}\backslash\mathcal{U},\mathcal{X}\backslash\mathcal{U}]}, {\tilde{C}^\dag}_{[\mathcal{E}_{\mathcal{XQ}},\mathcal{X}\backslash\mathcal{U}]})$ must have a minimum cost subgraph $\mathcal{P}^\dag(\mathcal{X}\cup\mathcal{Y}, \mathcal{E}_{\mathcal{P}^\dag},w_D')$, taking into account sensor to output link costs, which is spanned by an output cactus patch when any such node set $\mathcal{U}$ is removed.  Note that
			\begin{equation}\label{DesignProblem:MainPrf:Eq2}
				W(\mathcal{P}^\dag)	\leq W(\mathcal{D}(\tilde{A}^\dag,\tilde{C}^\dag))=F(\tilde{A}^\dag,\tilde{C}^\dag)
			\end{equation}
			where $W(\cdot)$ gives the total weight of a graph.  Furthermore there are at least $k+1$ internally node-disjoint paths beginning at $x$ and ending in $\mathcal{Y}$ in $\mathcal{P}^\dag$ for all $x\in\mathcal{X}$ because any graph disconnected from all outputs by $k$ sensor node failures could not have that property.  Reversing the process in Step~5, construct $\mathcal{T}^\dag(\mathcal{X}\cup\mathcal{Y}\cup\mathcal{Z}, \mathcal{E}_{\mathcal{T}^\dag},w_D')$ where $\mathcal{E}_{\mathcal{T}^\dag}=(\mathcal{E}_{\mathcal{P}^\dag} \backslash\{(x,x)|x\in\mathcal{X}\})\cup\{(y,z)|(x,y) \in\mathcal{E}_{\mathcal{P}^\dag}\textrm{ for some }x\in\mathcal{X}\}$.  Because the self-loops have zero cost and output links connecting to $z$ have weight $0$ with respect to $w_D'$, it follows that
			\begin{equation}\label{DesignProblem:MainPrf:Eq3}
				W(\mathcal{T}^\dag)=W(\mathcal{P}^\dag).
			\end{equation}
			Noting that $\kappa_{\mathcal{T}^\dag}(x,z)\geq k+1$ for all $x\in \mathcal{X}$, it is clear that $\mathcal{T}^\dag$ is a $z$-rooted $(k+1)$-node-connected Steiner subgraph of $\mathcal{G}_D'$ with terminal nodes $\mathcal{X}$.  By \eqref{DesignProblem:MainPrf:Eq1}-\eqref{DesignProblem:MainPrf:Eq3}
			\begin{equation}\label{DesignProblem:MainPrf:Eq4}
				W(\mathcal{T}^\dag)\leq F(\tilde{A}^\dag,\tilde{C}^\dag).
			\end{equation}
			Similarly,
			\begin{equation}\label{DesignProblem:MainPrf:Eq5}
				W(\mathcal{T}^*)=W(\mathcal{P}^*),
			\end{equation}
				and, this time with equality,
			\begin{equation}\label{DesignProblem:MainPrf:Eq6}
	W(\mathcal{P}^*)=W(\mathcal{D}(\tilde{A}^*,\tilde{C}^*))= F(\tilde{A}^*,\tilde{C}^*).
			\end{equation}
			Hence, by \eqref{DesignProblem:MainPrf:Eq5}-\eqref{DesignProblem:MainPrf:Eq6},		
			\begin{equation}\label{DesignProblem:MainPrf:Eq7}
				W(\mathcal{T}^*)=F(\tilde{A}^*,\tilde{C}^*),
			\end{equation}
			so it follows from \eqref{DesignProblem:MainPrf:Eq1} that
			\begin{equation}\label{DesignProblem:MainPrf:Eq8}
				W(\mathcal{T}^\dag)<W(\mathcal{T}^*).
			\end{equation}
			This contradicts the fact that $\mathcal{T}^*$ is a minimum $z$-rooted $(k+1)$-node-connected Steiner subgraph for $\mathcal{G}_D'$ with terminal nodes $\mathcal{X}$.  Thus, $(\tilde{A}^*,\tilde{C}^*)$ is a minimum cost solution to the optimization problem.  \hfill $\small \blacksquare$
		
		\end{proof}

		%***Existence condition verification***
\begin{remark}[Existence Verification and Maximum Robustness]\label{DesignProblem:ExistenceRmk} The condition that for each $x\in \mathcal{X}$ there is a set of at least $k+1$ internally node-disjoint directed paths in $\mathcal{G}_P$ each beginning at $x$ and ending in $\mathcal{Q}$ may be efficiently verified through $N=|\mathcal{X}|$ maximum $(\{x_s\},\mathcal{Q})$-flow computations (with restricted vertex capacity) where the source node is $x_s=x$, non-source nodes in $\mathcal{X}$ have capacity~1, and links in $\mathcal{E}_{\mathcal{XX}}\cup\mathcal{E}_{\mathcal{XQ}}$ have capacity~1.  This process can also be used to find the largest value of $k$ for which the solution exists.  Among all the nodes $x\in \mathcal{X}$ find the one with smallest node-capacitated flow to $\mathcal{Q}$.  The largest admissible value of $k$ is 1 less than that flow value.
\end{remark}
		
		Before applying Algorithm~1 to obtain an optimal dynamics structure $(\tilde{A}^*,\tilde{C}^*)$ that solves the problem in $\eqref{DesignProblem:V2:Eq1}-\eqref{DesignProblem:V2:Eq5}$, it would be prudent to ensure that a solution providing robustness factor $k$ exists for the input physical links cost graph $\mathcal{G}_P$.  In order to accomplish this, it must be verified that for each $x\in \mathcal{X}$ there is a set of at least $k+1$ internally node-disjoint paths beginning at $x$ and ending in $\mathcal{Q}$, in which case Theorem~\ref{DesignProblem:MainThm} guarantees that a solution exists.  This condition can be tested by computing the maximum flow from each $x_s$, treated as the only source, to $Q$, treated as a set of sinks, where all links $\mathcal{E}_{\mathcal{XX}}\cup\mathcal{E}_{\mathcal{XQ}}$ have capacity~1 and the nodes $x\in \mathcal{X}\backslash \{x_s\}$ also have capacity~1.  Through node splitting techniques \cite{FrankRootedKConnections}, the node capacities may be converted to link capacities.  By connecting all sinks to a common sink by links of infinite capacity, the result can then be formulated as a standard maximum flow problem, which can be solved in polynomial time by a number of algorithms, such as the Ford-Fulkerson algorithm \cite{CombinatorialOpt}.  Because each node can only be used once due to the node flow capacity, the desired condition holds if the maximum $(\{x_s\},\mathcal{Q})$-flow is at least $k+1$ for all $x_s\in \mathcal{X}$.
		
		%***Computational complexity***
		\begin{remark}[Implementation and Complexity]\label{DesignProblem:ComplexityRmk}
		
			The optimization in Step 4 may be simplified to a minimum $z$-rooted $k$-node-connected spanning subgraph computation by inserting an additional $k$ zero-cost parallel (duplicate) links from each $y\in\mathcal{Y}$ to $z$ in $\mathcal{G}_D'$.  This computation can be computed in polynomial time through maximum weighted matroid intersection methods detailed in \cite{FrankRootedKConnections}.
		
		\end{remark}

%		\begin{figure*}
%		
%		\begin{floatrow}
%		
%			\floatbox{figure}[\linewidth][\FBheight][t]
%			{}
%			{\centering
%			\includegraphics[width=\linewidth]{RobustFig.png}
%			\caption{The plot shows simulated network failure probability plotted against the fraction of sensor node failures for designed robustness levels $k=0,\ldots,3$ in networks with $|\mathcal{X}|=50$.  Results show a  gentler slope for higher values of $k$, even beyond the robustness guarantee.
%			}
%			\label{DesignProblem:RobustnessFig}}
%			\floatbox{figure}[\linewidth][\FBheight][t]
%			{}
%			{\centering
%			\includegraphics[width=.75\linewidth]{ExampleFig.png}
%			\caption{This illustrative example shows a small network on the unit square designed with $|X|=30$ randomly placed sensor nodes (black), and $|Q|=4$ backbone nodes (green), central node $Z$ (red), and robustness level $k=2$, and distance squared costs with maximum radius $R_{lim}=.4$.  The colors allow comparison to Figure~\ref{NetworkExampleA}.}
%		\label{DesignProblem:ExampleFig}}
%		
%		\end{floatrow}
%		
%	\end{figure*}
		
		While the optimization in Step 4 could be solved as a Steiner subgraph problem because all augmenting links have tail in the terminal nodes $\mathcal{X}$, it is simpler to convert the problem to a $z$-rooted $k$-node-connected spanning subgraph computation.  All nodes in $\mathcal{X}$ have $k+1$ internally node-disjoint paths to $Q$ assuming a solution exists.  Thus, each $x\in\mathcal{X}$ has $k+1$ internally node-disjoint paths to $\mathcal{Y}$, each ending at a different node. Nodes in $\mathcal{Y}$ have only one direct connection to $z$.  Therefore, addition of $k$ zero-cost parallel links from each $y\in \mathcal{Y}$ to $z$ makes $\mathcal{G}_D'$ $z$-rooted $k$-node-connected as a whole.  By computing the $z$-rooted $k$-node-connected spanning subgraph and discarding redundant zero-cost edges from $\mathcal{Y}$ to $\mathcal{Z}$, the solution is obtained.
		
		Note that Step~4 is, by far, the most costly step and subsumes the complexity of the other steps.  The $z$-rooted $k$-node-connected spanning subgraph can be computed in polynomial time as a maximum weighted common independent set of two matroids as described in \cite{FrankRootedKConnections}.  Matroids provide rules that define which subsets of a set are independent.  These rules must satisfy several axioms that will not be detailed in this paper.  For two matroids over edge set~$E$, the basic weighted matroid intersection algorithm involves $O(|S|^2||E|)$ calls to the two matroid independence oracles, functions that determine membership of  a set in the matroid \cite{CombinatorialOpt}, where $|S|$ is the maximum size of a common independent set.  In this case $|S|$, the maximum size of a common independent set, is $|S|=(k+1)(|\mathcal{X}|+|\mathcal{Y}|)$.  Links from $Y$ to $z$ are always included in the solution, so they can be trivially included in the final set, making $|S|$ is $O(k|\mathcal{X}|)$ for the purpose of counting calls to the matroid oracles.  The set is $E=\mathcal{E}_{\mathcal{XX}}\cup\mathcal{E}_{\mathcal{XY}} \cup\mathcal{E}_{\mathcal{YZ}}$, but links from $Y$ to $z$ are always included in the final solution, so $|E|$ is $O(|\mathcal{E}_{\mathcal{XX}}|+|\mathcal{E}_{\mathcal{XY}}|)$ for purposes of counting.  For this problem, one matroid oracle has relatively low complexity, while the other has complexity $O(|V||S|^2)$ \cite{FrankRootedKConnections} where $|V|$ is $O(|\mathcal{X}|+|\mathcal{Y}|)$. Thus the total complexity of the algorithm is $O(|V||E||S|^4)$ or $O(k^4|\mathcal{X}|^4(|\mathcal{X}|+|\mathcal{Y}|) (|\mathcal{E}_{\mathcal{XX}}|+|\mathcal{E}_{\mathcal{XY}}|))$.  It is claimed in \cite{FrankRootedKConnections} that the matroid oracle can be computed in $O(|V|^3)$, which would reduce this to $O(|V|^3|E||S|^2)$ or $O(k^2|\mathcal{X}|^2(|\mathcal{X}|+|\mathcal{Y}|)^3 (|\mathcal{E}_{\mathcal{XX}}|+|\mathcal{E}_{\mathcal{XY}}|))$.  Improved matroid intersection algorithms also exist.

		\begin{remark}[Special Case]\label{DesignProblem:SpecialCaseRmk} For the case in which $k=0$, the optimization in Step~4 reduces to a minimum $z$-rooted spanning arborescence of $\mathcal{G}_D'$ with terminal nodes $\mathcal{X}$.
		\end{remark}
		
		The special case $k=0$ of the minimum cost network design problem, which requires no robustness to node failures, was originally proposed and solved in \cite{PequitoKruzickEUSIPCO2013} when symmetric cost structure is considered.
 
 Under these conditions, the Steiner subgraph optimization step reduces to a minimum $z$-rooted Steiner arborescence problem on $\mathcal{G}_D'$ with terminal nodes $\mathcal{X}$.  Because each $y\in \mathcal{Y}$ may be automatically connected directly to~$z$ by the non-augmenting link $(y,z)$, this can be further reduced to a minimum spanning tree arborescence, leading to Remark~\ref{DesignProblem:SpecialCaseRmk}.  Thus, this case may be computed with greater simplicity using Edmond's algorithm \cite{PequitoKruzickEUSIPCO2013}.
				
		%***Practical robustness statistics***
		\begin{remark}[Observable Instantiations]\label{DesignProblem:InstantiationsRmk} An observable instantiation $(A,C)$ of an optimal structurally observable solution $(\tilde{A}^*,\tilde{C}^*)$ output by Algorithm~1 over a finite field $\mathbb{F}_{p^n}$ can be found randomly with high probability for fields of large order $p^n$.  As $p^n$ grows without bound, the probability of an observable random instantiation over $\mathbb{F}_{p^n}$ approaches~1 \cite{SundaramHadjicostis2}.
		\end{remark}
		
		Although the output of the algorithm gives an optimal structure $(\tilde{A}^*,\tilde{C}^*)$ of the system dynamics that satisfies the constraints, an observable instantiation $(A,C)$ of the dynamics must be obtained for any implementation.  If the field in which the system operates is $\mathbb{F}=\mathbb{R}$ or $\mathbb{F}=\mathbb{C}$, such instantiations of the structure are guaranteed to exist and the set of unobservable realizations over those fields has zero measure.  Consequently, a random instantiation of the structure from a suitable distribution would be almost surely observable with probability one.

		While interesting, this approach suffers from the possibility of producing systems with poorly conditioned observability matrices that are full rank yet present numerical problems.  Additionally, physical devices cannot truly operate with general real or complex numbers.  These problems may both be avoided through the use of finite fields $\mathbb{F}=\mathbb{F}_{p^n}$ where $p$ is prime and $n$ is a positive integer. It has been shown in~\cite{SundaramHadjicostis2} that for an output tree with self-loops attached to the $N$ state nodes, an observable instantiation is guaranteed to exist over $\mathbb{F}_{p^n}$ if $p^n\geq N$.  In fact, Corollary 1 of~\cite{SundaramHadjicostis2} shows that this instantiation can be produced by assigning a distinct field element to each self-loop and the multiplicative identity to each link between distinct nodes.  Instead, if elements are chosen randomly at each sensor, repetition can be avoided and observability achieved with high probability if the field is sufficiently large, as shown in Theorem 5 of~\cite{SundaramHadjicostis2}, approaching probability~1 as $p^n$ increases without bound.  This provides a more advantageous approach to practical implementation and is summarized by Remark~\ref{DesignProblem:InstantiationsRmk}.
		
		\begin{figure}
		
			\centering
			\includegraphics[width=\linewidth]{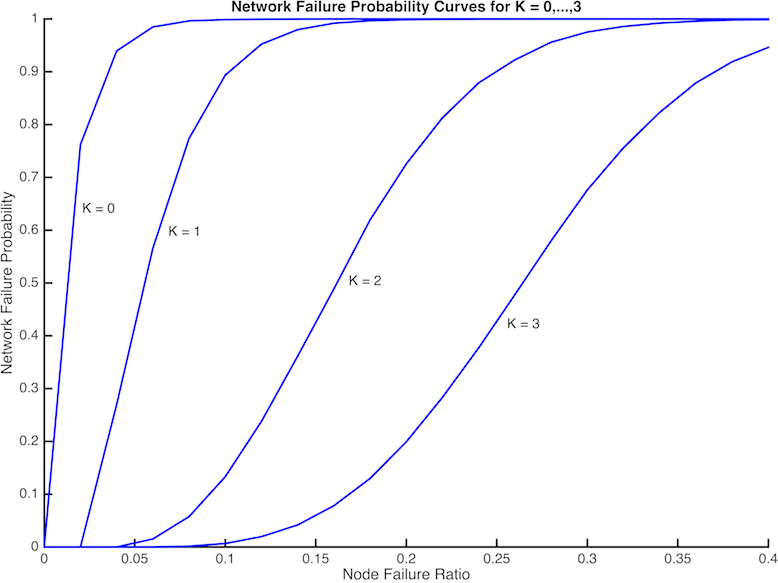}
			\caption{The plot shows simulated network failure probability plotted against the fraction of sensor node failures for designed robustness levels $k=0,\ldots,3$ in networks with $|\mathcal{X}|=50$.  Results show a  gentler slope for higher values of $k$, even beyond the robustness guarantee.}
		\label{DesignProblem:RobustnessFig}
		\end{figure}
		
		\begin{figure}
		
			\centering
			\includegraphics[width=.75\linewidth]{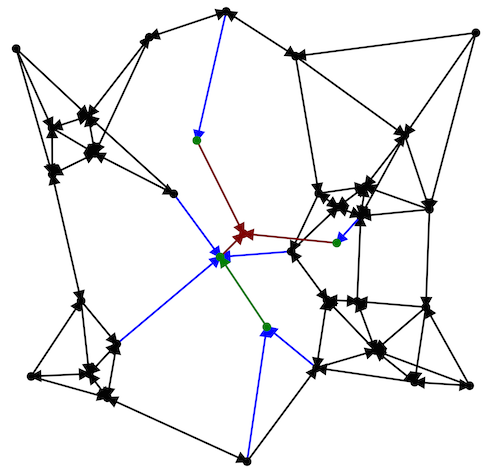}
			\caption{This illustrative example shows a small network on the unit square designed with $|X|=30$ randomly placed sensor nodes (black), and $|Q|=4$ backbone nodes (green), central node $Z$ (red), and robustness level $k=2$, and distance squared costs with maximum radius $R_{lim}=.4$.  The colors allow comparison to Figure~\ref{NetworkExampleA}.}
		\label{DesignProblem:ExampleFig}
		\end{figure}
		\begin{remark}[Empirical Robustness]\label{DesignProblem:RobustnessRmk} While Algorithm~1 guarantees robustness to failure of any node subset $\mathcal{U}\subset \mathcal{X}$ with $|\mathcal{U}|\leq k$, the solution may be robust in practice to a larger number of randomly chosen node failures.  Here, as in the rest of the paper, robustness refers to the fact that at least one spanning output cactus remains intact after $k$ node failures.  That is, no surviving nodes are disconnected and, thus, unobservable.  For more than $k$ node failures, the idea of robustness can be relaxed from this guarantee to examine the failure probability.
		\end{remark}
		%
%		\begin{figure}[t]
%			\centering
%			\includegraphics[width=\linewidth]{RobustFig.png}
%			\caption{The plot shows simulated network failure probability plotted against the fraction of sensor node failures for designed robustness levels $k=0,\ldots,3$ in networks with $|\mathcal{X}|=50$.  Results show a  gentler slope for higher values of $k$, even beyond the robustness guarantee.}
%			\label{DesignProblem:RobustnessFig}
%		\end{figure}
		%\newpage
		In practice, a network designed by Algorithm~1 may be robust to more sensor node failures than the guaranteed degree of robustness $k$.  Although the algorithm guarantees that removal of any subset of failing sensors $\mathcal{U}\subseteq\mathcal{X}$ with $|\mathcal{U}|\leq k$  does not break the structural observability of the remaining network, it does not necessarily follow that the network fails if $|\mathcal{U}|=\ell>k$.  For instance, the structure of the sensor subnetwork for $k=0$ is a $Q$ rooted branching with loops, and failure of any leaf node does not affect the remainder of the network.  This effect becomes more pronounced for higher values of $k$.  With $\ell$ node failures selected uniformly at random, the probability of network failure cannot be easily computed.  However, it is possible to empirically simulate it for networks constructed from pseudo-randomly placed nodes with pseudo-random node failures.
		
Figure \ref{DesignProblem:RobustnessFig} plots the empirical probability of network failure for pseudo-randomly generated node locations against the node failure ratio $\ell/|\mathcal{X}|$ for designed degree of robustness $k=0,...,3$.  This simulation used $|\mathcal{X}|=50$ uniformly distributed sensor nodes, $|\mathcal{Y}|=3$ backbone nodes, and distance squared link cost.  Figure \ref{DesignProblem:ExampleFig} represents the graph of a smaller network designed in such a way, with the additional constraint on the communication radius.  For each randomly generated network and failure ratio value, $\ell$ sensor nodes were randomly selected to fail.  The network fails if any surviving node no longer has a directed path to a backbone node.  The expected network failure probability was computed for each node failure ratio over 100 random graphs each with 1000 random sets of failing nodes.  As seen in Figure \ref{DesignProblem:RobustnessFig}, the resulting networks can have low failure probability even beyond the robustness guarantee, where failure occurs with probability zero.  The plot also demonstrates that when designing for higher guaranteed robustness level, the failure probability grows much more slowly.  For instance, when designing for robustness level $k=3$, the failure probability is only approximately $0.2$ when $10$ nodes out of $50$ sensor nodes fail.
	\section{Conclusions}\label{Conclusions}
The distributed networks of agents or sensors examined in this paper model situations in which nodes update their state variables using values obtained by sensing state information from nodes within their local neighborhoods.  The state of some nodes is directly sensed by nodes in a network backbone, which are directly accessed by a central fusion center. The fusion center computes the (global) network state from the data it obtains.  This type of scenario may arise in robotics formation control applications or in field inversion problems, for instance.  By duality of observability and controllability, the same technique may be used to drive the sensor states to some desired target.
	
	This paper posed the problem of finding minimum cost network update dynamics with respect to a link cost objective function under an observability constraint.  The polynomial time algorithm proposed finds the optimal network that is structurally observable when any $k$ sensor nodes are removed by using combinatorial optimization algorithms, guaranteeing the existence of observable dynamics that can be found separately in a straightforward way by random instantiation over sufficiently large finite fields.  In practice, networks may survive more node failures than the guaranteed amount, especially as $k$ increases.  This paper improves the results presented in~\cite{PequitoKruzickEUSIPCO2013}, extending consideration to robustness requirements and designable network backbone topologies.  Future efforts could address a number of interesting network design problems, such as finding optimal networks that are simultaneously both structurally observable and structurally controllable.  Alternate problem formulations could decrease network iterations by attempting to reduce the observability index. %, increasing the potential communication rate.
 Additional work could also focus on implementing the network design process in a fully distributed manner within the provided framework, using known distributed algorithms for computing minimum spanning arborescences in the simple $k=0$ case.
	
	%***Bibliography section***
	\bibliographystyle{IEEEtran}
	\bibliography{StructurallyObservableDistributedNetworks}
	
\end{document}